\documentclass[10pt,twocolumn,showpacs,preprint,amsmath,amssymb,aps,prd,eqsecnum]{revtex4}
\usepackage{amsmath,amssymb}
\usepackage{graphicx}
\usepackage{indentfirst} 
\usepackage{tabularx}
\begin{document}

\title{Formation of naked singularities in five-dimensional space-time}

\author{Yuta Yamada$^{1}$
\footnote{yamada@is.oit.ac.jp} and 
Hisa-aki Shinkai$^{1, 2}$
\footnote{shinkai@is.oit.ac.jp}}
\address{$^{1}$~Faculty of Information Science and Technology, Osaka Institute of Technology, 1-79-1 Kitayama, Hirakata, Osaka 573-0196, Japan\\
$^{2}$~Computational Astrophysics Laboratory, Institute of Physical \& Chemical Research (RIKEN), Hirosawa, Wako, Saitama 351-0198, Japan~ }

\date{\today}

\begin{abstract}
We numerically investigate the gravitational collapse of collisionless particles
in spheroidal configurations both in four and five-dimensional (5D) space-time. 
We repeat the simulation performed by Shapiro and Teukolsky (1991) that announced
an appearance of a naked singularity, and also find that the similar results 
in 5D version. 
That is, in a collapse of a highly prolate spindle, the Kretschmann invariant blows up
outside the matter and no apparent horizon forms.  
We also find that the collapses in 5D proceed rapidly than in 4D, and 
the critical prolateness for
appearance of apparent horizon in 5D is loosened compared to 4D cases. 
We also show how collapses differ with spatial symmetries comparing 5D
evolutions in single-axisymmetry, SO(3), and those in double-axisymmetry, 
U(1)$\times$U(1).
\end{abstract}

\maketitle
\clearpage

\section{Introduction}
The so-called ``large extra-dimensional models" as a consequence of 
brane-world pictures have changed 
our viewpoints for a way of understanding the fundamental forces.
The scenarios of unifying the gravity at TeV scale or so open the
possibility of verification of higher-dimensional space-time models
at the CERN Large Hadron Collider (LHC). 
If the LHC detects productions (and evaporations) of mini 
black-holes as expected, 
then humankind will encounter a 
Copernican change of our outlook of the Universe.

With this background, black-holes in higher dimensional space-time are
extensively studied for a decade. 
Many interesting discoveries of new solutions have been reported, and 
their properties are being revealed.  However, 
fully relativistic dynamical features, such as the formation processes,
stabilities and late-time fate, are still unknown and 
they are waiting to be studied. 
Several groups including us begin reporting numerical studies 
in various topics in higher-dimensional models\cite{yamada_shinkai, yoshinoshibata, shibatayoshino, espanol_group, espanol_group2, pretorius}.

We, in this article, report our numerical simulations on 
gravitational collapse in axisymmetric space-time. 
The topic has been studied in many ways in $(3+1)$-dimensional space-time (4D, hereafter), 
among them the most impressive result we think is the work by Shapiro and Teukolsky
\cite{ST91} (ST91, hereafter); a highly prolate matter collapse
which may form a naked singularity.  
We repeat their simulations and also compare them with $(4+1)$-dimensional (5D) versions. 

In classical general relativity, it is well known that a space-time singularity will be generally formed in gravitational collapse of non-singular asymptotically flat initial data.
If a singularity forms without an event horizon, all physical predictions become invalid.  
In order not to occur such a disastrous situation, Penrose proposed the {\it cosmic censorship conjecture} \cite{penrose69}, which states that singularities are always clothed by event horizons.  

On the other hand, for nonspherical gravitational collapses, Thorne proposed 
the {\it hoop conjecture} \cite{thorne72} which states that black holes with horizons are formed when and only when a mass gets compacted into a small region.  
He expressed the compactness with a {\it `hoop'} around matter. 
If matter configuration is highly aspherical, then the hoop length becomes larger.
If so, the conjectured inequality does not hold, i.e. horizon will not be formed.
Thereby, the hoop conjecture indicates that a highly aspherical matter collapse 
will lead to a naked singularity. 

ST91 numerically showed that axisymmetric space-time with 
collisionless matter particles in spheroidal distribution will collapse to singularity, and there are no apparent horizon formed when the spheroids are highly prolate. 
The behaviors are consistent with their initial data analysis \cite{nakamura88}, and support the hoop conjecture. 
However, since numerical evolutions cannot provide the final structures nor the conclusive information for formation of naked singularities, debates were raised after their announcement. 
For example, Wald {\it et al} \cite{wald1,wald2} showed examples that 3-dimensional 
hypersurfaces can hit singularities nevertheless the space-time is consistent with cosmic censor. 
Their examples are not directly related with the numerical results with ST91, but we learned that a numerical result provides only a limited evidence. 

Regarding to the 5D cases, the hoop conjecture is supposed to be  
replaced with the {\it hyper-hoop} version\cite{ida,yoo,BFL04,senovilla0709,gibbons0903}, 
i.e. a criteria is not a hoop but a surface.  
In our previous work\cite{yamada_shinkai}, we numerically constructed 
initial data sequences of non-rotating matter for 5D evolutions and examined 
the hyper-hoop conjecture using minimum {\it area} around the matter. 
We found that the areal criteria matches with the appearance of apparent horizons
for spindle matter configurations (but not for ring configurations).
The sequences suggest that a highly prolate spindle in 5D will form a naked singularity similar to the 4D cases.  
We also found that a condition for a naked singularity formation is relaxed in 5D than 4D cases.  

One of the objectives of the present work is to compare the dynamics 
between 4D and 5D.  A simple estimation from the free-falling time indicates
that the gravitational collapse in 5D takes longer time than 4D cases.  
We show this is not applicable to the highly non-linear final stages. 
In 5D, two axes can be settled as rotational symmetric axes, so that we also compare
gravitational collapses in axisymmetry with those in ``doubly"-axisymmetric space-time. 

\section{Numerical code}
\begin{figure}[b]
\centering
\includegraphics[width=7.0cm, clip]{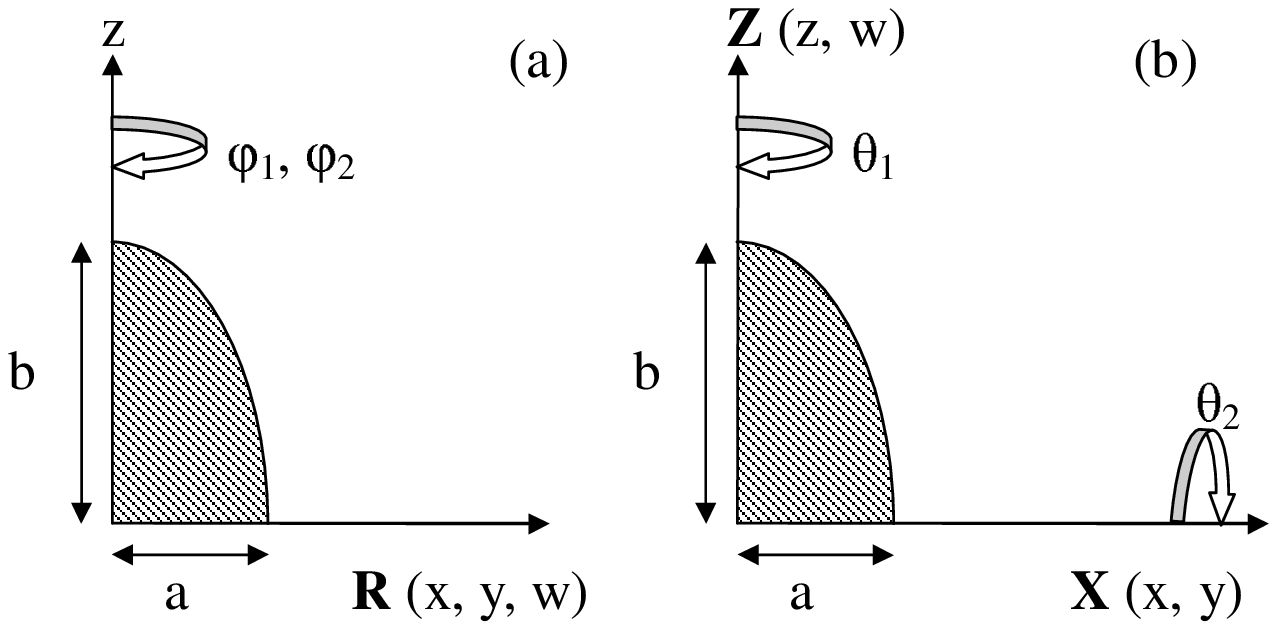}
\caption{\label{fig1}We evolve five-dimensional (a) axisymmetric [SO(3)] or (b)
double-axisymmetric [U(1)$\times$ U(1)], asymptotically flat space-time using the Cartesian
grid. The initial matter configuration is expressed with parameters $a$ and $b$. }
\end{figure}

We evolve five-dimensional axisymmetric [symmetric on $z$-axis, SO(3)] or 
doubly-axisymmetric [symmetric both on $x$ and $z$-axes, U(1)$\times$U(1)], 
asymptotically flat space-time (see Figure \ref{fig1}).  For the comparison, we also 
performed four-dimensional axisymmetric space-time evolutions.

We start our simulation from time symmetric 
and conformally flat initial data, 
which are obtained by solving the Hamiltonian constraint equations \cite{yamada_shinkai}. 
The asymptotical flatness is imposed throughout the evolution, which settles
the fall-off condition to the metric as $\sim 1/r$ for 4D cases
and $\sim 1/r^2$ for 5D cases. 

The matter is described with 5000 collisionless particles, which 
move along the geodesic equations.  We smooth out the matter 
by expressing each particle with Gaussian density distribution function
with its typical width is twice as much as the numerical grid. 
The particles are homogeneously distributed in a spheroidal shape, 
parametrized with $a$ and $b$ (Figure \ref{fig1}), or eccentricity $e=\sqrt{1-a^2/b^2}$.  

By imposing axisymmetry or double-axisymmetry, 
our model becomes practically a ($2+1$)-dimensional problem. 
We construct our numerical grids with the Cartesian coordinate $(x, z)$, and 
apply the so-called Cartoon method \cite{cartoon,yoshinoshibata} to 
recover the symmetry of space-time. 

The space-time is evolved using the Arnowitt-Deser-Misner (ADM) evolution equations. 
It is known that the ADM evolution equations excite an unstable mode 
(constraint-violation mode) in long-term simulations \cite{shinkai2009, shinkaiyoneda2004}.  However, we are free from this problem since 
gravitational collapse occurs within quite short time.  
By monitoring the violation of constraint equations during evolutions,
we confirm that our numerical code has second-order convergence, and also 
that the simulation continues in stable manner.  
The results shown in this report are obtained with numerical grids, 
129$\times$129$\times$2$\times$2.  We confirmed that higher resolution runs do not change the physical results. 

We use the maximal slicing condition for the lapse function $\alpha$, and
the minimal strain condition for the shift vectors $\beta^i$. 
Both conditions are proposed for avoiding the singularity
in numerical evolutions \cite{smarryork78}, 
and the behavior of $\alpha$ and $\beta^i$ roughly 
indicates the strength of gravity, conversely. 
The iterative Crank-Nicolson method is used for integrating ADM evolution equations, 
and the Runge-Kutta method is used for matter evolution equations. 

For discussing physics, we search the location of apparent horizon (AH), 
calculate the Kretschmann invariant (${\cal I}=R_{abcd}R^{abcd}$) 
on the spacial hypersurface.

\section{Results}
\begin{figure}[htbp]
  \begin{center}
    \begin{tabular}{cc}
          \resizebox{40mm}{!}{\includegraphics{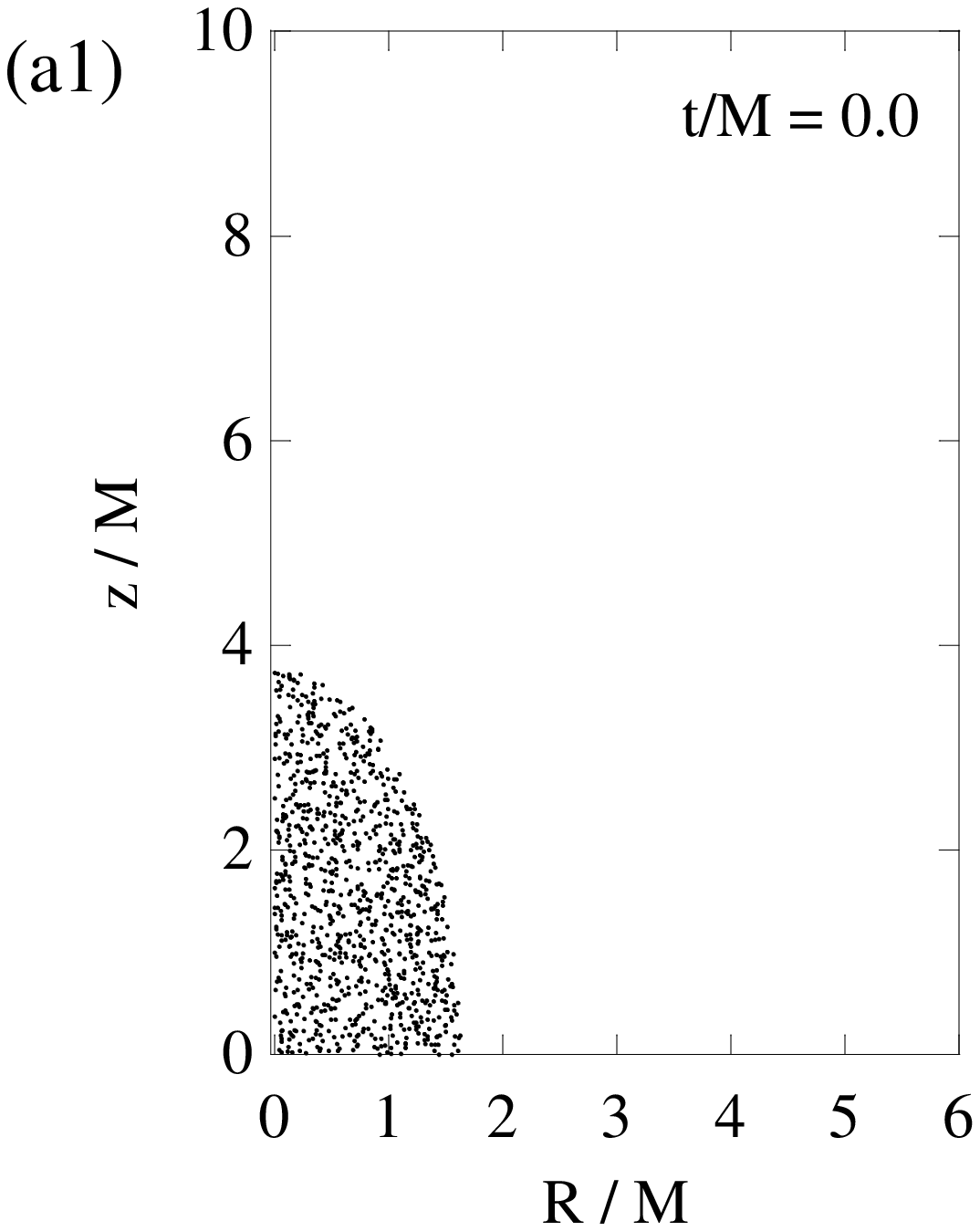}} &
      \resizebox{40mm}{!}{\includegraphics{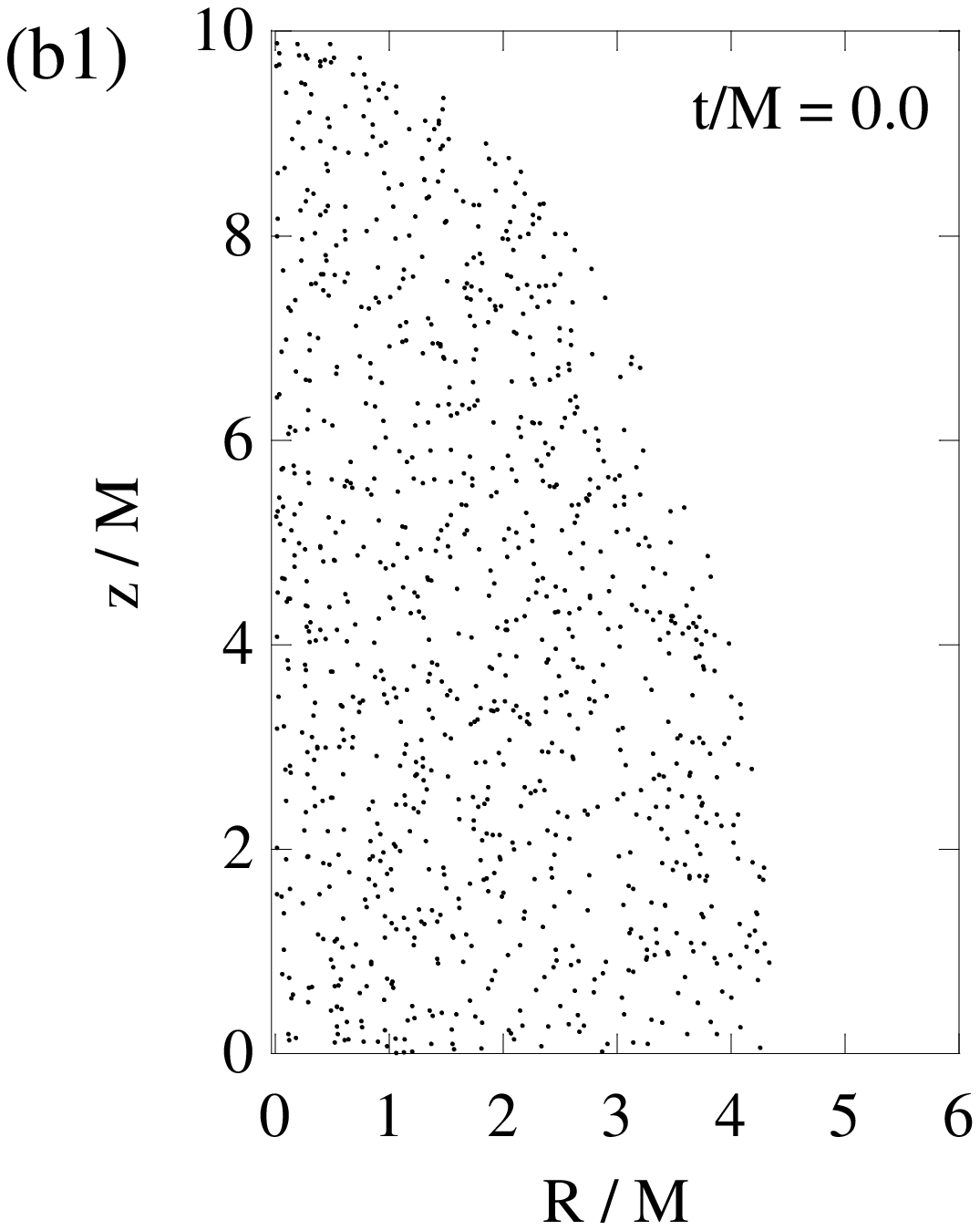}}
    \end{tabular}
  \end{center}
  \begin{center}
    \begin{tabular}{cc}
      \resizebox{40mm}{!}{\includegraphics{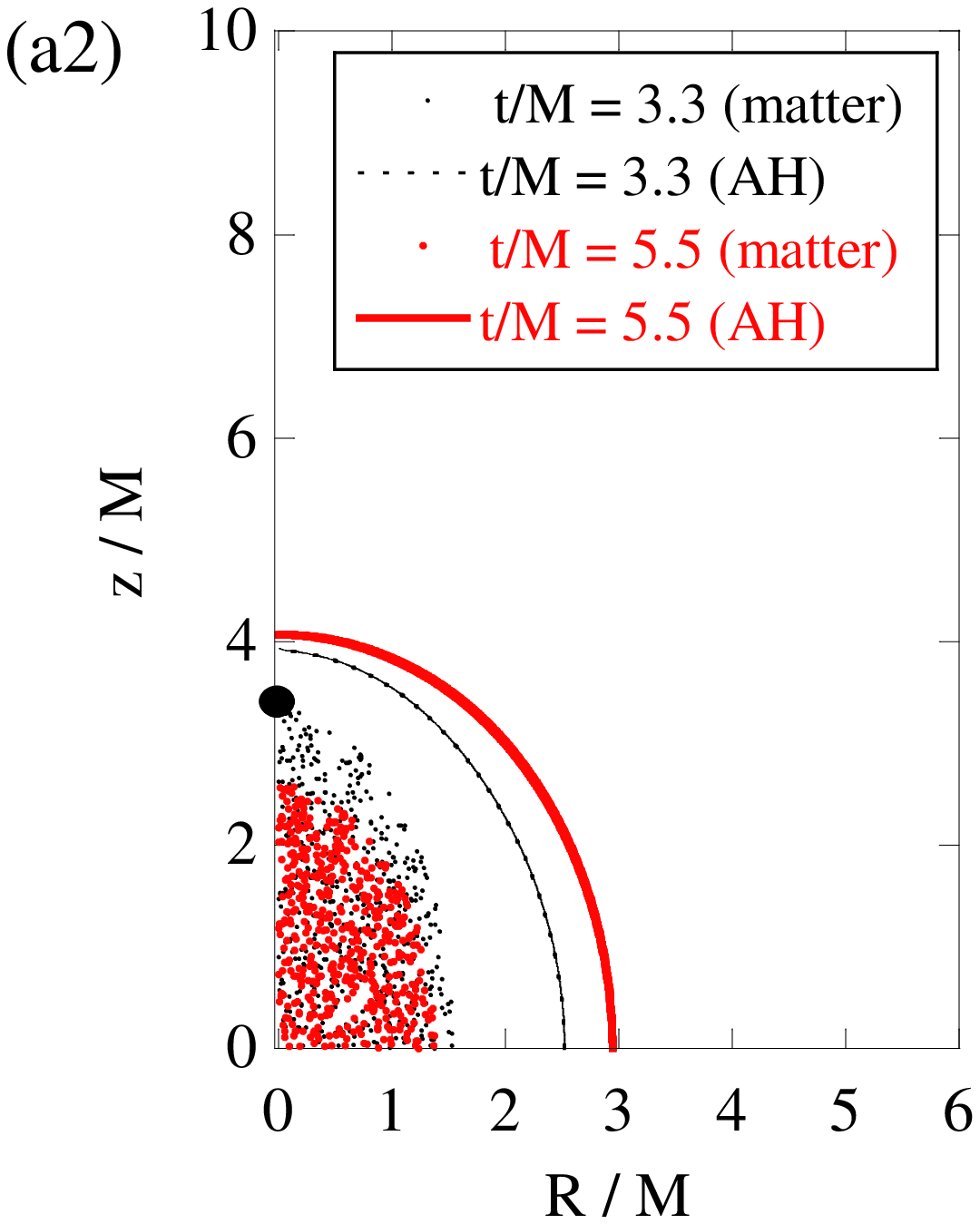}} &
      \resizebox{40mm}{!}{\includegraphics{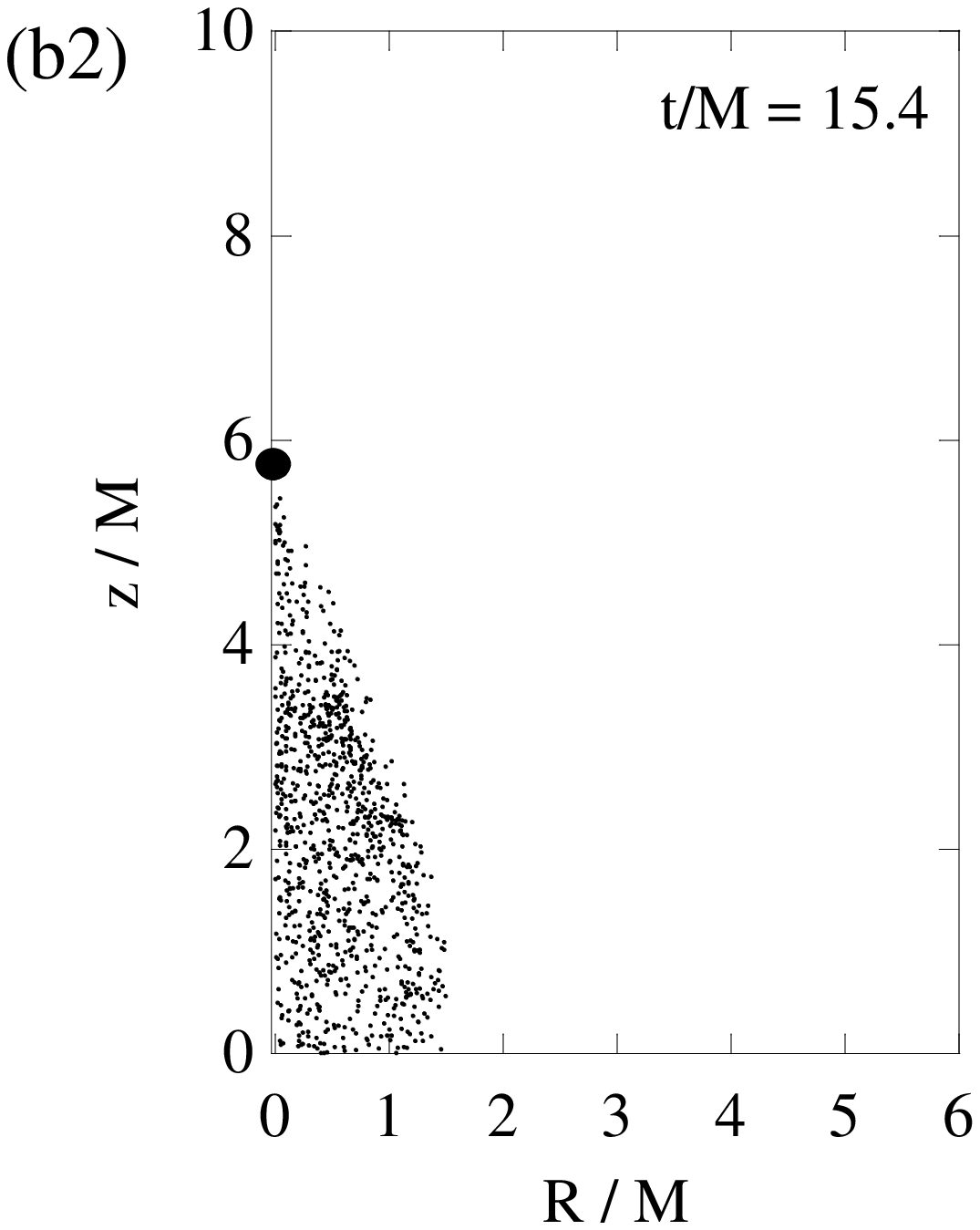}}
    \end{tabular}
  \end{center}
    \caption{\label{snapshot}
Snapshots of 5D axisymmetric evolution with the initial matter 
distribution of $b/M=4$
 [Fig.(a1) and (a2); model {\tt 5DS$\beta$} in Table \ref{table1}] 
and $10$ [Fig.(b1) and (b2); model {\tt 5DS$\delta$}].
We see 
the apparent horizon (AH) is formed at the coordinate time 
$t/M = 3.3$ for the former model 
and the area of AH increases, 
while AH is not observed for the latter model up 
to the time $t/M = 15.4$ when our code stops due to the large curvature.  
The big circle indicates the location of the maximum Kretschmann invariant 
${\cal I}_{\rm max}$ at the final time at each evolution. 
Number of particles are reduced to 1/10 for figures.        
}
\end{figure}

We prepare several initial data fixing the total ADM mass 
and the eccentricity of distribution, $e=0.9$.  By changing 
the initial matter distribution sizes, we observe the different 
final structures.  Figure \ref{snapshot} shows snapshots of 
5D axisymmetric evolutions of model $b/M=4$ and 10 
(model {\tt 5DS$\beta$} and  {\tt 5DS$\delta$}, respectively; see Table \ref{table1});   
the former collapses to a black hole while the latter collapses without AH formation. 

\begin{figure}[htbp]
\centering
\includegraphics[width=6cm, clip]{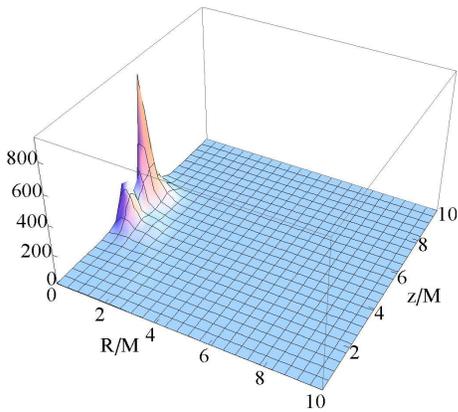}
\caption{\label{3Dplot}
Kretschmann invariant ${\cal I}$ for model {\tt 5DS$\delta$} at $t/M = 15.4$. 
The maximum is $O(1000)$, and its location is on $z$-axis, just outside of the matter. 
}
\end{figure}

All the models we tried result in forming a singularity (i.e., diverging ${\cal I}$). 
We stopped our numerical evolutions when the shift vector was not obtained with 
sufficient accuracy due to the large curvature.  For model {\tt 5DS$\delta$}, we integrated up to the coordinate time 
$t/M = 15.4$ and the maximum of the Kretschmann invariant ${\cal I}_{\rm{max}}$ 
became $O(1000)$ on $z$-axis (see Figure \ref{3Dplot}), but AH was not formed. 

\begin{table}[tbh] 
  \begin{center}
    \begin{tabular}{c|cccc}
    \hline 
    \hline 
    $b/M~(t=0)$ & 2.50 & 4.00 & 6.25 & 10.00 \\
    \hline 
    \hline 
    4D axisym. & {\tt 4D$\alpha$} & {\tt 4D$\beta$}  & {\tt 4D$\gamma$}  & {\tt 4D$\delta$}  \\
     & AH-formed & no  & no  & no  \\
     & $e_{\rm AH}=0.90$ &  &  &  \\
     & $e_{\rm f}=0.92$ & $e_{\rm f}=0.89$ & $e_{\rm f}=0.92$  & $e_{\rm f}=0.96$  \\
               \hline
    5D axisym. & {\tt 5DS$\alpha$} & {\tt 5DS$\beta$}  & {\tt 5DS$\gamma$}  & {\tt 5DS$\delta$}  \\
   SO(3)  & AH-formed & AH-formed  & no  & no  \\
     & $e_{\rm AH}=0.88$ & $e_{\rm AH}=0.88$  &  &  \\
     & $e_{\rm f}=0.82$ & $e_{\rm f}=0.84$  & $e_{\rm f}=0.88$  & $e_{\rm f}=0.96$  \\
               \hline
    5D double   & {\tt 5DU$\alpha$} & {\tt 5DU$\beta$}  & {\tt 5DU$\gamma$}  & {\tt 5DU$\delta$}  \\
    ~~ axisym.& AH-formed & AH-formed  & AH-formed  & no  \\
   U(1)$\times$U(1)  & $e_{\rm AH}=0.86$ & $e_{\rm AH}=0.87$  & $e_{\rm AH}=0.92$  &  \\
   & $e_{\rm f}=0.79$ & $e_{\rm f}=0.81$  & $e_{\rm f}=0.90$  & $e_{\rm f}=0.98$  \\
\hline
    \end{tabular}
  \end{center}
    \caption{\label{table1} Model-names and the results of their evolutions whether we observed AH or not.  The eccentricity $e$ of the collapsed matter configurations is also shown; 
$e_{\rm AH}$ and $e_{\rm f}$ are at the time of AH formation (if formed), and on the numerically obtained final hypersurface, respectively.
}
\end{table}

When the initial matter is highly prolate, AH is not observed.
This is consistent with 4D cases \cite{nakamura88,ST91}, and matches with 
the predictions from initial data analysis in 5D cases \cite{yoo,yamada_shinkai}.
The location of ${\cal I}_{\rm{max}}$ is on $z$-axis, and just outside of the matter\footnote{The Kretschmann invariant expresses the strength of the curvature, which is determined by the gradient of metric. For example, when we solve a single star with uniform density, the maximum value of the metric appears at the center of matter configuration, but the maximum value of the metric gradient appears off-center and likely at the outside of matter region. Therefore, our results of the location of the maximum Kretschmann invariant is not so odd.}.
This is again the same with 4D cases \cite{ST91}. 
The absence of AH with diverging ${\cal I}$ suggests a formation of 
naked singularity in 5D.

\begin{figure*}[htbp]
  \begin{center}
    \begin{tabular}{ccc}
          \resizebox{50mm}{!}{\includegraphics{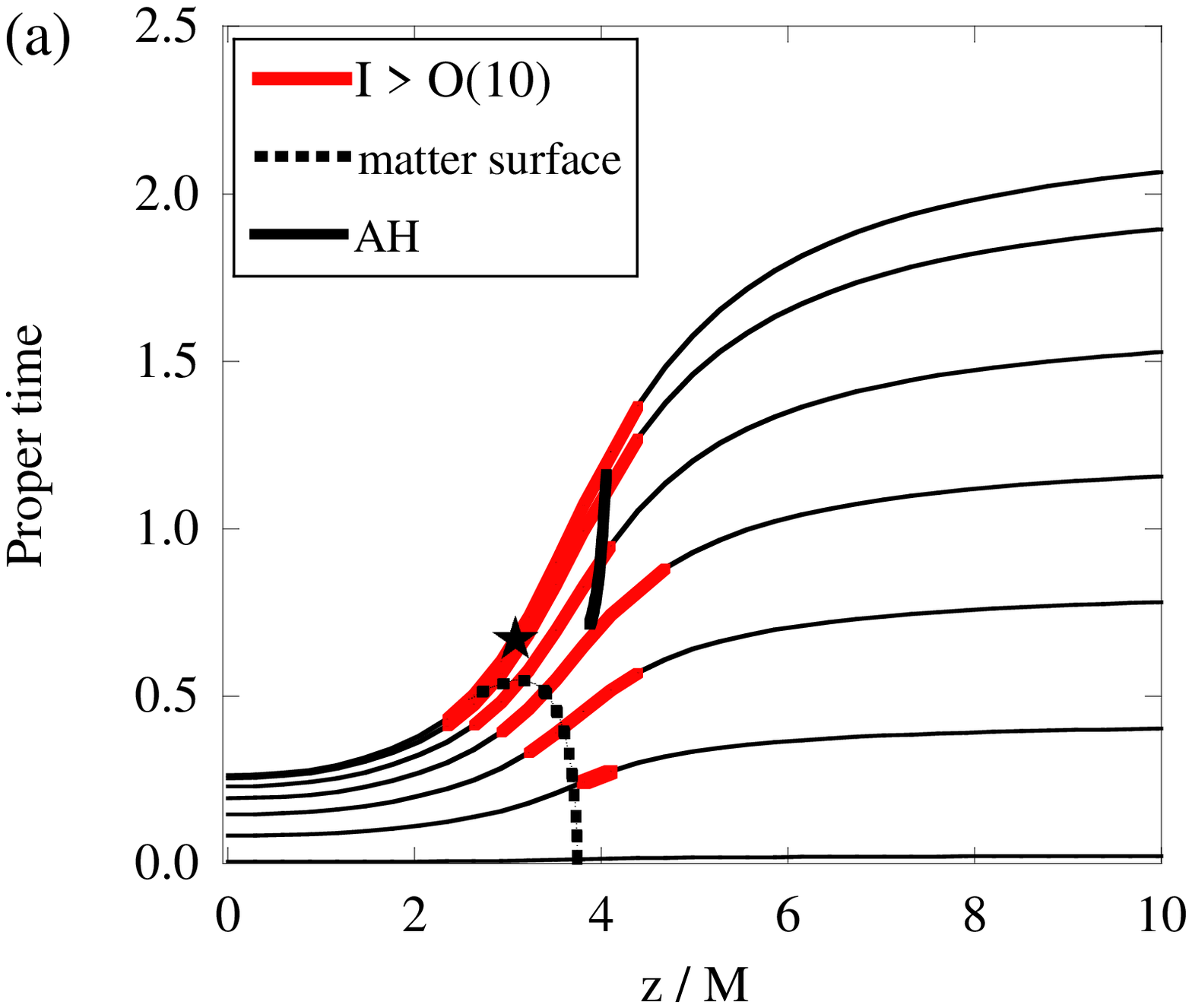}} \;\;\;\;\;\;&
      \resizebox{50mm}{!}{\includegraphics{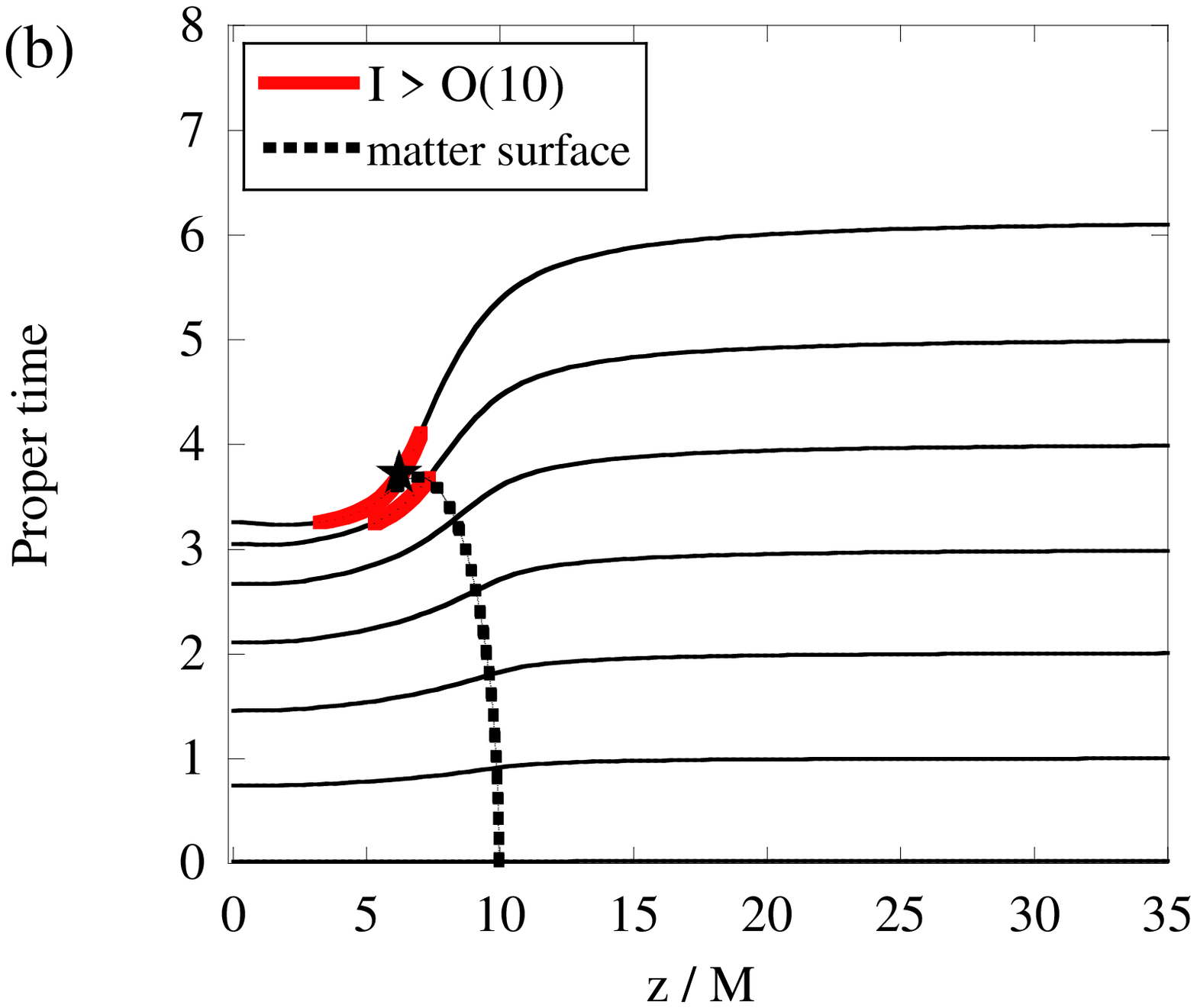}} \;\;\;\;\;\;&
      \resizebox{50mm}{!}{\includegraphics{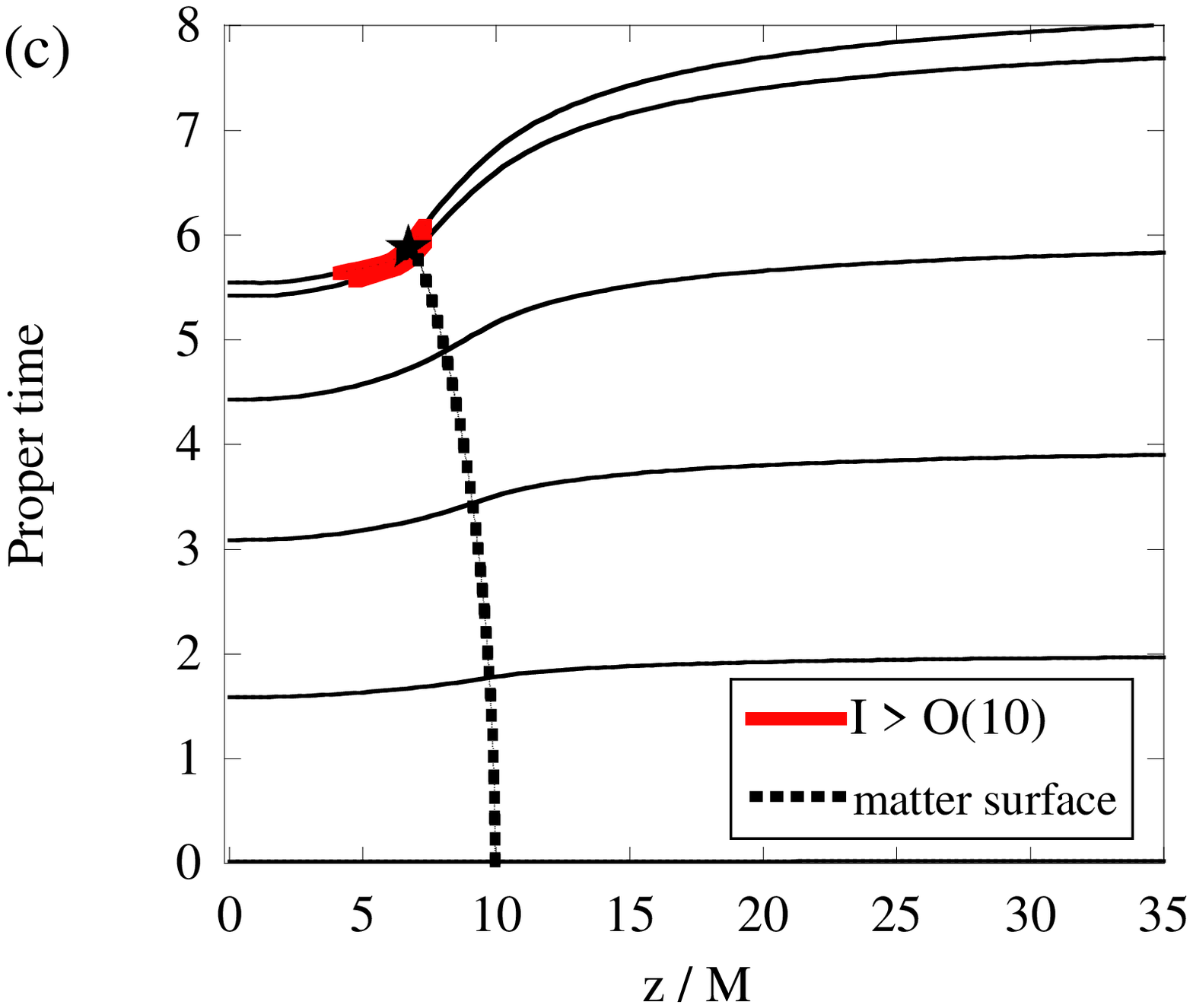}}\\
    \end{tabular}
  \end{center}
    \caption{\label{proper_time}
The snapshots of the hypersurfaces on the $z$-axis in the proper-time versus 
coordinate diagram; 
(a) model {\tt 5DS$\beta$}, (b) model {\tt 5DS$\delta$}, and (c) model {\tt 4D$\delta$}. 
The upper most hypersurface is the final data in numerical evolution. 
We also mark the matter surface and the location of AH if exist.
The ranges with ${\cal I}\geq 10$ are marked with bold lines and peak value 
of ${\cal I}$ express by asterisks. }
\end{figure*}

In order to compare the results with 4D and 5D, we reproduced the 
results of ST91. 
We then find that the $e=0.9$ initial data with $b/M=10$ (model {\tt 4D$\delta$}) collapses 
without forming AH, and the code stops at the coordinate time $t=20.91$ 
with ${\cal I}_{\rm max}=84.3$ on the $z$-axis ($z/M=6.1$); 
all the numbers match quite well with ST91.  (Note that our slicing conditions
and coordinate structure is not the same with ST91.) 

Figure \ref{proper_time} is for the comparisons of
hypersurfaces for the 5D models which collapses 
(a) with forming AH, (b) without forming AH, and (c)
4D collapses without forming AH.  
We see hypersurfaces are bending due to the slicing conditions, 
and figures tell us how numerically integrated region covers the space-time. 

We also performed 5D collapses with doubly-axisymmetric [U(1)$\times$U(1)] space-time.
The matter and space-time evolve quite similar to 5D and 4D axisymmetric cases, 
but we find that the critical configurations for forming AH is different. 
Table \ref{table1} summarizes the main results of 4D and two 5D cases. 
We find that AH in 5D is formed in larger $b$ initial data than 4D cases. 
This result is consistent with our prediction 
from the sequence of initial data \cite{yamada_shinkai}. 
AH criteria with initial $b$ is loosened for 5D doubly-axisymmetric 
cases.

We also show the eccentricity of matter, $e_{\rm AH}$ and $e_{\rm f}$, 
at the time of AH formation (if formed) and at the final time in the simulation, 
respectively. 
The numbers in Table \ref{table1} indicate that the eccentricity itself is not a guiding measure for AH formation, but they give us a hint for understanding the differences. 

In 4D, the eccentricity increases after AH is formed ({\tt 4D$\alpha$}), while
it decreases in 5D axisymmetric cases ({\tt 5DS$\alpha$} and {\tt 5DS$\beta$}). 
That is, the 5D collapses proceed towards spherical configurations. 
This fact would be explained by the degree of freedom of the movements. 
In general, 5D spacetime is expected to produce much gravitational radiation than 4D
spacetime \cite{shibatayoshino, espanol_group2}, 
since there are more modes of oscillation exist.  
Gravitational radiation normally works to 
change shapes to spherical, because it is produced from the acceleration of spacetime
and carries the energy away. (It is known that compact binary system will evolve into 
a circular orbit due to the emission of gravitational radiation.)  
Therefore collapses in 5D spacetime are likely to evolve towards more spherical. 
This interpretation together with the hoop conjecture will explain 
why AH-formation condition is loosened in 5D cases. 

In the 5D doubly-axisymmetric cases, on the other hand, the magnitude of 
$e_{\rm f}$ is smaller than 5D axisymmetric cases for small $b/M$ cases 
({\tt 5DS$\alpha$} vs. {\tt 5DU$\alpha$}, or {\tt 5DS$\beta$} vs. {\tt 5DU$\beta$}), 
while it is larger for large $b/M$ cases 
({\tt 5DS$\gamma$} vs. {\tt 5DU$\gamma$}, or {\tt 5DS$\delta$} vs. {\tt 5DU$\delta$}). 
We think this is because that
the doubly-axisymmetric collapses proceed in more symmetric manner 
than axisymmetric collapse near the origin, while they proceed in 
more 4D-like axisymmetric collapses near the axes far from 
the origin.  The collapses of small $b/M$ initial data, therefore, 
will evolve into more spherical shape, 
while the large $b/M$ initial data will evolve increasing the eccentricity, where 
the latter is similar to 4D cases.

\begin{figure}[bpht]
\centering
\includegraphics[width=7cm, clip]{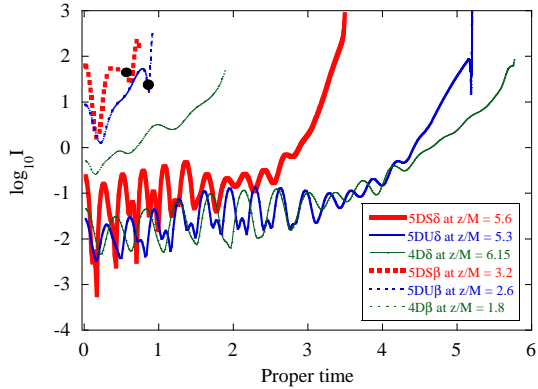}
\caption{\label{max_kretschmann} 
The Kretschmann invariant ${\cal I}$ at the location of ${\cal I}_{\rm{max}}$ on the
final hypersurface is plotted as a function of proper time at its location.
Labels indicate model-names in Table \ref{table1}.
The time of AH formation (t=0.6 for model {\tt 5DS$\beta$}, t=0.9 for  {\tt 5DU$\beta$}) is shown by a dot.
}
\end{figure}

In Figure \ref{max_kretschmann}, 
we plot ${\cal I}$ at the point which gives ${\cal I}_{\rm max}$ on the final 
hypersurface as a function of proper time. 
The ${\cal I}$ diverges at the end of simulations in all the cases, but
the diverging time becomes later for larger $b/M$ initial data. 
We see that 5D-collapses are generally proceeding more rapidly than 4D collapses.
We also see that collapses in 5D doubly-axisymmetric space-time is 
proceeding more slowly than 5D single axisymmetric cases.
If we observe further, the model {\tt 5DU$\beta$} evolves quite similar to 
{\tt 5DS$\beta$}, while  {\tt 5DU$\delta$} evolves quite similar to 
{\tt 4D$\delta$}.  These behaviors support the previous understandings of
the evolution of the eccentricity.

\section{Discussions}
In this article, we reported our numerical study of gravitational collapses
in 5D space-time.   
We collapsed spheroidal matter expressing with collisionless particles, 
and observed the evolution of the Kretschmann invariant and the apparent horizon (AH) formation. 

The collapsing behaviors are generally quite similar to the cases in 4D, 
but we also found that 
(a) 5D-collapses proceed rapidly than 4D-collapses, 
(b) AH appears in highly prolate matter configurations than 4D cases, 
(c) doubly-axisymmetric [U(1)$\times$U(1)] assumption makes collapse proceed
towards more spherical when it forms AH, but presents quite similar behavior 
with 4D cases for large configurations, and 
(d) the positive evidence for appearance of a naked
singularity in 5D. 

Up to this moment, we only checked the existence of apparent horizons, and 
not the event horizons. The system does not include any angular momentum. 
We are implementing our code to cover these studies.  

We are now preparing our next 
detail report including the validity of 
hyper-hoop conjecture in 5D, and the cases of the ring objects. 

We thank T. Torii and B. K. Tippett for discussion. 
This work was supported partially by the Grant-in-Aid for
Scientific Research Fund of Japan Society of the Promotion of Science,
No. 22540293.
Numerical computations were carried out on Altix3700 BX2 at YITP in Kyoto University,
and on the RIKEN Integrated Cluster of Clusters (RICC).

\end{document}